\documentclass{ws-procs9x6}

\begin{document}

\title{Dynamical Low-Mass Fermion Generation in Randall-Sundrum Background
\footnote{\uppercase{T}he main part of this paper is based on the works in 
collaboration with \uppercase{K}.~\uppercase{F}ukazawa, 
\uppercase{Y}.~\uppercase{K}atsuki, \uppercase{T}.~\uppercase{M}uta and 
\uppercase{K}.~\uppercase{O}hkura.${}^7$}}

\author{T.~Inagaki}

\address{Information Media Center, Hiroshima University,\\
Higashi-Hiroshima, 739-8521, Japan\\ 
E-mail: inagaki@hiroshima-u.ac.jp}

\maketitle

\abstracts{
It is investigated that a dynamical mechanism to generate a low mass fermion 
in Randall-Sundrum (RS) background. We consider a five-dimensional 
four-fermion interaction model with two kinds of bulk fermion fields and take 
all the mass scale in the five-dimensional spacetime at the Planck scale. 
Evaluating the effective potential of the induced four-dimensional model, 
I calculate the dynamically generated fermion mass. It is shown that 
dynamical symmetry breaking takes place and one of the fermion mass is 
generated at the electroweak scale in four dimensions.
}

\section{Introduction}
To construct a unified theory of the electroweak interaction, strong 
interaction and gravity it is important to make investigation on the gauge 
hierarchy problem, how the electroweak scale is realized in the theory at 
the Planck scale.
As in a large extra-dimension model it is possible to solve the gauge 
hierarchy problem to consider a four-dimensional brane embedded in a 
higher-dimensional spacetime.\cite{Antoniadis:90ew,Arkani-Hamed:98rs} 
Randall and Sundrum considered a higher-dimensional curved spacetime with 
negative curvature and found a beautiful solution of the hierarchy problem 
by using the exponential factor in the metric.\cite{Randall:99ee} 
Here we launched a plan to study a dynamical mechanism to realize the 
electroweak scale from the Planck scale physics in a model of the brane 
world proposed by 
Randall and Sundrum.\cite{Abe:01yb,Abe:01yi,Abe:02yb,prepare:02}

\begin{figure}[ht]
\centerline{\epsfxsize=3.9in\epsfbox{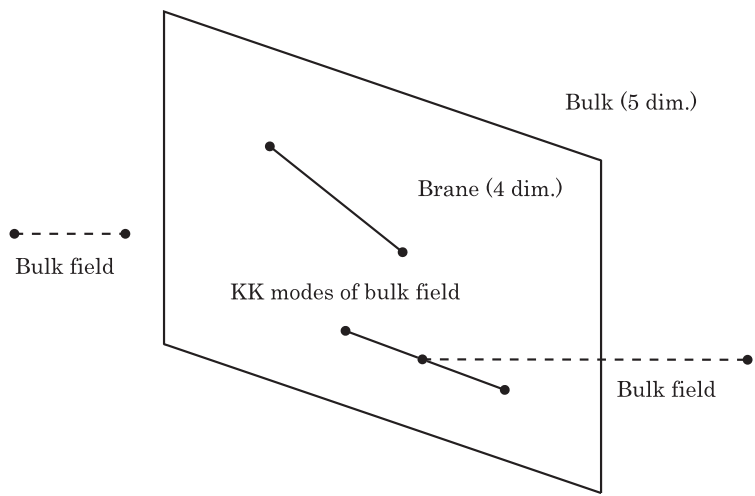}}   
\caption{Image of the brane world. \label{inagaki-fig1}}
\end{figure}

At the beginning, it is considered that only the graviton can propagate 
in the extra-dimension and all the standard model particles are localized 
on the four-dimensional brane. However, there is a possibility that some of 
the standard model particles also propagate in the 
extra-dimension.\cite{Chang:99nh} In Fig.~\ref{inagaki-fig1} we illustrate 
an image of a four-dimensional brane embedded in a five-dimensional 
bulk. The bulk fields are the fields which can propagate in the bulk. 
The KK excitation modes of the bulk fields appear on the brane and the 
modes may affect some of low energy phenomena.
 
One of the interesting phenomena is found in a spontaneous electroweak 
symmetry breaking. The electroweak symmetry can be dynamically broken 
down due to the fermion and anti-fermion condensation.
Many works have been done to see the contribution of the KK modes to 
dynamical symmetry breaking in models with large extra 
dimensions.\cite{Ishikawa:uu,Dobrescu:98dg,Cheng:99bg,Abe:00ny,Arkani-Hamed:00hv,Hashimoto:00uk,Gusynin:02cu} 
Here a theory with bulk fermions is considered in the RS background.
We assume the existence of two types of bulk fermion fields which can 
propagate in a five-dimensional balk. To construct a model where the 
fermion field naturally develops the electroweak mass scale, a four-fermion 
interaction is introduced between these bulk fermions. As is known, 
the four-fermion interaction model is a simple model of dynamical symmetry 
breaking. It is expected that a negative curvature enhances symmetry 
breaking.\cite{Inagaki:93ya,Inagaki:95bk,Inagaki:97kz,Rius:01dd} 
Evaluating the induced four-dimensional effective potential, we calculate 
the mass scale of the fermion in four dimensions.  Since we are interested 
in the bulk standard model particles, the KK excitations of graviton are 
assumed to have no serious effect on the fermion mass and ignore them.

\section{Four-Fermion Interaction Model in Randall-Sundrum Background}
Here we briefly review the Randall-Sundrum idea\cite{Randall:99ee} and 
introduce a four-fermion interaction between bulk fermions. 

\subsection{Randall-Sundrum Background}\label{subsec:RS}
\begin{figure}[ht]
\centerline{\epsfxsize=3.2in\epsfbox{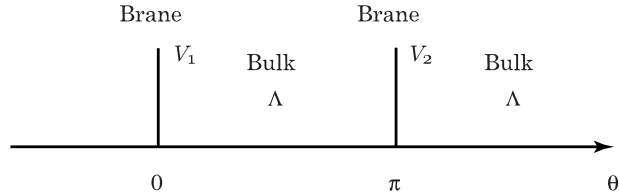}}
\caption{Randall-Sundrum spacetime \label{inagaki-fig3}}
\end{figure}

The RS background is a five-dimensional spacetime whose fifth dimension is 
compactified on an orbifold with $S^1/Z_2$ symmetry and two Minkowski 
branes exist at the orbifold fixed points, $\theta=0$ and $\pi$, 
see Fig.~\ref{inagaki-fig3}.
The background spacetime is a static solution of the Einstein equation 
if the cosmological constant in the bulk, $\Lambda$, and that on the 
brane, $V_1, V_2$, satisfy the relationship,
\begin{equation}
\Lambda = -V_1 = V_2.
\label{cond:RS}
\end{equation}
The spacetime described by the metric,
\begin{equation}
g^{\mu\nu}=e^{-2kr|\theta|}\eta_{\mu\nu}dx^{\mu}dx^{\nu}+r^2d\theta^2 .
\label{metric}
\end{equation}
It is a maximally symmetric spacetime with a constant negative curvature, 
i.e., five-dimensional anti de-Sitter spacetime. 

The warp factor $e^{-2kr|\theta|}$ in Eq.(\ref{metric}) plays an important 
role to solve the hierarchy problem. The effective Planck scale, mass scale 
for gravity on the brane, is given by
\begin{equation}
{M_{pl}}^2\sim\frac{M^3}{k}(1-e^{-2kr\pi})\sim\frac{M^3}{k} ,
\end{equation}
where $M$ is the fundamental scale in the bulk and $k$ is the curvature.
On the other hand, the mass scale for $M_{phys}$ on the $\theta=\pi$ brane
is suppressed by the warp factor.
\begin{equation}
M_{phys}=M e^{-kr\pi} .
\end{equation}

For $k\sim 11$, the electroweak mass scale, $M_{EW}$, can be realized 
from only the Planck scale, $M_{pl}$, without introducing some large number. 
\begin{equation}
\left\{
\begin{array}{l}
M\sim k\sim \mbox{O}(M_{pl}) , \\
M_{phys}\sim \mbox{O}(M_{EW}) .
\end{array}
\right.
\end{equation}
This is the most important mechanism of the RS model. We want to realize 
this mechanism dynamically and construct a model where the fermion mass 
is generated at the electroweak scale.

\subsection{Bulk Four-Fermion Interaction Model}\label{subsec:FF}
We study the bulk four-fermion interaction model defined by
\begin{equation}
{\mathcal L}^{5D}=\sqrt{-G}\left[
\bar{\psi}_1^{5D}i\partial\!\!\!/ \psi_1^{5D}
+\bar{\psi}_2^{5D}i\partial\!\!\!/ \psi_2^{5D}
+\lambda(\bar{\psi}_1^{5D}\psi_2^{5D})(\bar{\psi}_2^{5D}\psi_1^{5D})
\right] ,
\label{L:ff}
\end{equation}
where we assume the existence of two kinds of bulk fermions with different
parity,
\begin{equation}
\left\{
\begin{array}{l}
\psi_1^{5D}(x,\theta)= \gamma_5 \psi_1^{5D}(x,-\theta) ,\\
\psi_2^{5D}(x,\theta)= -\gamma_5 \psi_2^{5D}(x,-\theta) .
\end{array}
\right.
\end{equation}
Two kinds of fermion necessary for constructing Dirac mass term on the brane. 
It is possible to consider the other types of four-fermion interaction, 
for example $(\bar{\psi}_1^{5D}\psi_1^{5D})(\bar{\psi}_1^{5D}\psi_1^{5D})$ 
and $(\bar{\psi}_2^{5D}\psi_2^{5D})(\bar{\psi}_2^{5D}\psi_2^{5D})$. But 
the interaction in Eq.(\ref{L:ff}) is essential to generate a low mass mode.

Following the procedure in Ref. 8 we derive the mode expansion of the bulk 
fermion in the RS background. 
\begin{equation}
\psi^{5D}(x,\theta)=\sum_{n=0}^{\infty}\psi_{R}^{(n)}(x)g_{R}^{(n)}(\theta)
+\psi_{L}^{(n)}(x)g_{L}^{(n)}(\theta) ,
\label{KK:psi}
\end{equation}
where $g_{L}^{(n)}$ and $g_{R}^{(n)}$ are left and right mode functions
which satisfy
\begin{equation}
\left\{
\begin{array}{l}
\displaystyle \int d\theta e^{-3kr|\theta|} g_{L}^{(n)}(\theta)
g_{L}^{(n)}(\theta) = \delta_{mn} , \\[1.4mm]
\displaystyle \int d\theta e^{-3kr|\theta|} g_{R}^{(n)}(\theta)
g_{R}^{(n)}(\theta) = \delta_{mn} .
\end{array}
\right.
\end{equation}

In practical calculation it is more convenient to introduce auxiliary field 
$\sigma\sim\bar{\psi}_1\psi_2$. Applying the KK mode expansions 
(\ref{KK:psi}), the Lagrangian (\ref{L:ff}) reads
\begin{eqnarray}
&&{\mathcal L}=\bar{\psi}^{(0)}_{1R}i\gamma^{\mu}\partial_{\mu}\psi^{(0)}_{1R}
+\bar{\psi}^{(0)}_{2L}i\gamma^{\mu}\partial_{\mu}\psi^{(0)}_{2L}
\nonumber \\
&&+\sum_{1\leq n}\left[
\bar{\psi}^{(n)}_{1R}i\gamma^{\mu}\partial_{\mu}\psi^{(n)}_{1R}
+\bar{\psi}^{(n)}_{1L}i\gamma^{\mu}\partial_{\mu}\psi^{(n)}_{1L}
+\bar{\psi}^{(n)}_{2R}i\gamma^{\mu}\partial_{\mu}\psi^{(n)}_{2R}
+\bar{\psi}^{(n)}_{2L}i\gamma^{\mu}\partial_{\mu}\psi^{(n)}_{2L}
\right]
\nonumber \\
&&+\sum_{m,n=0}^{\infty}
\left(\bar{\psi}^{(m)}_{1R}\ \bar{\psi}^{(m)}_{2R}\
\bar{\psi}^{(m)}_{1L}\ \bar{\psi}^{(m)}_{2L}\right)
M
\left(
\begin{array}{c}
\psi^{(n)}_{1R} \\
\psi^{(n)}_{2R} \\
\psi^{(n)}_{1L} \\
\psi^{(n)}_{2L}
\end{array}
\right)
-\int d\theta r \sqrt{-G} \frac{|\sigma|^2}{\lambda} .
\label{L:KK}
\end{eqnarray}
$M$ corresponds to the fermion mass. It is a function of the vacuum 
expectation value of $\sigma$ and the mode functions.\cite{prepare:02}

\section{Dynamically Generated Fermion Mass}
To obtain the fermion mass in four dimensions we need to calculate the 
vacuum expectation value of $\sigma$ which is determined by observing 
the minimum of the induced four-dimensional effective potential.
Integrating over the extra direction in Eq. (\ref{L:KK}), we obtain the 
induced four-dimensional theory. Since the RS background has no 
translational invariance along to the extra direction it is impossible 
to generally perform the integration over the extra direction. 
Here we restrict ourselves in some specific forms of the vacuum 
expectation value, $\langle\sigma\rangle = v e^{kr\theta}$ 
and $\langle\sigma\rangle = v$ where $v$ is a constant parameter.

\begin{figure}[b]
\centerline{\epsfxsize=3.4in\epsfbox{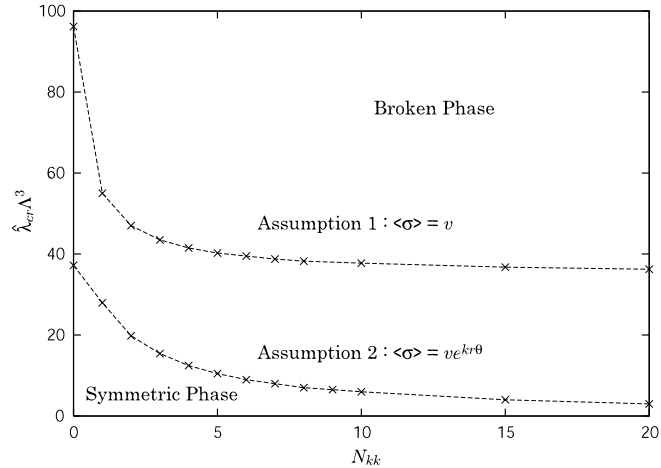}}   
\caption{Critical coupling constant as a function of the truncated 
scale $N_{kk}$. \label{inagaki-fig2}}
\end{figure}
After some numerical calculations we obtain the behaviors of the effective 
potential in both the cases and find the critical coupling where the 
vacuum expectation value disappears.\cite{prepare:02} 
In Fig.~\ref{inagaki-fig2} we draw the behavior of the critical coupling. 
$\hat{\lambda}$ is defined by 
$\hat{\lambda}\equiv (1-e^{-4k r\pi}) \lambda/(4k)$ and $N_{kk}$ is a 
truncated scale of KK mode summations. It is natural to take 
$N_{kk}\sim O(10^{16})$.
In the region between two critical lines the state, 
$\langle\sigma\rangle = v e^{kr\theta}$, is more stable than 
$\theta$-independent one.
For a large $N_{kk}$ limit the critical coupling is proportional to 
$1/N_{kk}$ in the $\langle\sigma\rangle = v e^{kr\theta}$ case. 
But the critical coupling seems to be a constant, 
$\hat{\lambda_{cr}}\sim O(30)/\Lambda^3$, in the $\theta$-independent case. 
We conclude that the natural scale of the four-fermion coupling, 
$\hat{\lambda}\sim{1/M_{pl}^3}$, is located between two critical lines 
and $\theta$-dependent vacuum is realized. 

For $\langle\sigma\rangle = v e^{kr\theta}$ the fermion mass matrix of 
the induced four-dimensional theory reduces to
\begin{equation}
M = 
\left(
\begin{array}{cc}
m_n & -v \\
-v  & m_n
\end{array}
\right) ,
\label{mass}
\end{equation}
where $m_n$ is given by
\begin{equation}
m_n = \frac{nk\pi}{e^{k\pi r}-1}, \ \
n=\cdots, -2, -1, 0, 1, 2, \cdots .
\end{equation}
The eigen values of the mass matrix is described as
\begin{equation}
m_f \sim \left| v + \frac{nk\pi}{e^{k\pi r}-1} \right| , \ \
n=\cdots, -2, -1, 0, 1, 2, \cdots .
\end{equation}
It corresponds to the mass for each KK modes in four dimensions.
We can choose $n$ where $m_f$ is smaller than 
$k\pi/(e^{k\pi r}-1) \sim M_{EW}$. There is a mode whose mass is smaller 
than the electroweak scale,  $M_{EW}$, even if $v$ develops a value 
near the Planck scale. Therefore a low mass fermion is generated dynamically 
in the bulk four-fermion interaction model.

\section{Conclusion}
The dynamical origin of the electroweak mass scale have been investigated in 
the RS background. We have assumed the existence of two kinds of bulk fermion 
fields with different parity and studied a bulk four-fermion interaction 
model. Evaluating the effective potential for two specific 
$\theta$-dependence of the state, we have calculated the critical value of 
the four-fermion coupling and found the more stable state. In a natural 
choice of all the physical parameters the vacuum expectation value depends 
on the extra direction. In the stable state the fermion mass term has been 
analytically calculated. We have shown the existence of a mode whose mass 
is smaller than the electroweak scale. The electroweak mass scale can be 
realized from only the Planck scale in the RS brane world due to the fermion 
and anti-fermion condensation. This is one of the dynamical realizations of 
the so-called Randall-Sundrum mechanism.

There are some remaining problems. We consider only two specific forms of 
the vacuum state and conclude the state whose expectation value of 
$\sigma$ has the form $v e^{kr\theta}$ is more stable. To find the true 
vacuum we must calculate the induced effective potential for a general 
form of $\langle\sigma\rangle$.

The fermion and anti-fermion condensation may affect the structure of 
spacetime. To analyze the spacetime evolution the behavior of the 
stress tensor is under investigation.

\section*{Acknowledgments}
The author would like to thank H.~Abe, K.~Fukazawa, Y.~Katsuki, T.~Muta 
and K.~Ohkura for stimulating discussions.


\begin{thebibliography}{0}
\bibitem{Antoniadis:90ew}
I.~Antoniadis,
{\it Phys. Lett.\/} {\bf B 246}, 377 (1990). 

\bibitem{Arkani-Hamed:98rs}
N.~Arkani-Hamed, S.~Dimopoulos and G.~R.~Dvali,
{\it Phys. Lett.\/} {\bf B 429}, 263 (1998). 

\bibitem{Randall:99ee}
L.~Randall and R.~Sundrum,
{\it Phys. Rev. Lett\/} {\bf 83}, 3370 (1999). 

\bibitem{Abe:01yb}
H.~Abe, T.~Inagaki and T.~Muta, {\it in:\/} ``Fluctuating Paths and Fields '', ed. W. Janke, A. Pelster, H.-J. Schmidt, M. Bachmann, World Scientific, 2001.

\bibitem{Abe:01yi}
H.~Abe, K.~Fukazawa and T.~Inagaki,
{\it Prog. Theor. Phys.\/} {\bf 107}, 1047 (2002). 

\bibitem{Abe:02yb}
H.~Abe and T.~Inagaki,
{\it Phys. Rev.\/} {\bf D 66}, 085001 (2002). 

\bibitem{prepare:02} 
K.~Fukazawa, T.~Inagaki, Y.~Katsuki, T.~Muta and K.~Ohkura,
hep-ph/0308022.

\bibitem{Chang:99nh}
S.~Chang, J.~Hisano, H.~Nakano, N.~Okada and M.~Yamaguchi,
{\it Phys. Rev.\/} {\bf D 62}, 084025 (2000). 

\bibitem{Ishikawa:uu}
K.~Ishikawa, T.~Inagaki, K.~Yamamoto and K.~Fukazawa,
{\it Prog. Theor. Phys.\/} {\bf 99}, 237 (1998). 

\bibitem{Dobrescu:98dg}
B.~A.~Dobrescu,
{\it Phys. Lett.\/} {\bf B 461}, 99 (1999). 

\bibitem{Cheng:99bg}
H.~C.~Cheng, B.~A.~Dobrescu and C.~T.~Hill,
{\it Nucl. Phys.\/} {\bf B 589}, 249 (1998). 

\bibitem{Abe:00ny}
H.~Abe, H.~Miguchi and T.~Muta,
{\it Mod. Phys. Lett.\/} {\bf A 15}, 445 (2000). 

\bibitem{Arkani-Hamed:00hv}
N.~Arkani-Hamed, H.~C.~Cheng, B.~A.~Dobrescu and L.~J.~Hall,
{\it Phys. Rev.\/} {\bf D 62}, 096006 (2000). 

\bibitem{Hashimoto:00uk}
M.~Hashimoto, M.~Tanabashi and K.~Yamawaki,
{\it Phys. Rev.\/} {\bf D 64}, 056003 (2001). 

\bibitem{Gusynin:02cu}
V.~Gusynin, M.~Hashimoto, M.~Tanabashi and K.~Yamawaki,
{\it Phys. Rev.\/} {\bf D 65}, 116008 (2002). 

\bibitem{Inagaki:93ya}
T.~Inagaki, T.~Muta and S.~D.~Odintsov,
{\it Mod. Phys. Lett.\/} {\bf A 8}, 2117 (1993). 

\bibitem{Inagaki:95bk}
T.~Inagaki,
{\it Int. J. Mod. Phys.\/} {\bf A 11}, 4561 (1996). 

\bibitem{Inagaki:97kz}
T.~Inagaki, T.~Muta and S.~D.~Odintsov,
{\it Prog. Theor. Phys. Suppl.\/} {\bf 127}, 93 (1997). 

\bibitem{Rius:01dd}
N.~Rius and V.~Sanz,
{\it Phys. Rev.\/} {\bf D 64}, 075006 (2001). 

\end{thebibliography}
\end{document}